\documentclass[printer]{aa}

\usepackage{txfonts}
\usepackage[dvips]{graphicx}
\usepackage{natbib}

\def\EQ{\begin{equation}}
\def\EN{\end{equation}}
\def\EQA{\begin{eqnarray}}
\def\ENA{\end{eqnarray}}
\def\uu{{\bf u}}
\def\vv{{\bf v}}

\def\UU{{\bf U}}

\def\OOmega{{\mbox{\boldmath$\Omega$}}}

\begin{document}

\title{Multi-scale theory of rotating turbulence}

\author{Nicolas Leprovost \and Eun-jin Kim}

\institute{Department of Applied Mathematics, University of Sheffield, Sheffield S3 7RH, UK}

\date{Received / Accepted }

\abstract{}{To understand the dynamics of stellar interiors, we study the effect of rotation on turbulence.}{We consider turbulence induced by an arbitrary forcing and derive turbulence amplitude and turbulent transport coefficients (turbulent viscosity and diffusivity), first by using a quasi-linear theory and then by using a multi-scale renormalisation analysis.}{With an isotropic forcing, the quasi-linear theory gives that the turbulent transport coefficients, both parallel and perpendicular to the rotation vector, have the asymptotic scaling $\Omega^{-1}$ for rapid rotation (i.e. when the rotation rate $\Omega$ is larger than the inverse of the correlation time of the forcing and the diffusion time), while the renormalisation analysis suggests a weaker dependence on $\Omega$, with $\Omega^{-1/2}$ scaling. The turbulence amplitude is found to scale as $\Omega^0 - \Omega^{-1}$ in the rapid rotation limit depending on the property of the forcing. In the case of an anisotropic forcing with inhibited motion in the vertical direction, as should be relevant in a strongly stratified medium, we find that non-diffusive fluxes of angular momentum scale as $\Omega^{-2} - \Omega^{-1}$ for rapid rotation, depending on the temporal correlation of the forcing. We discuss the implications of our result for the dynamics of stellar interiors.}{}


\titlerunning{Multi-scale theory of rotating turbulence}

\maketitle

\section{Introduction}
Rotation and turbulence are ubiquitous features of many astrophysical bodies and play a crucial role in the dynamics of these bodies. For instance, turbulence is very often invoked as a means to increase the transport of angular momentum [e.g. the accretion rate in disks \citep{Balbus98}] or the mixing of light elements in stars [e.g. to explain the surface depletion \citep[see][and references therein]{ Pinsonneault97}]. Since the rotation rate of stars varies with their mass (spectral types) and age, it is important to develop a general theory of turbulent transport which is valid for arbitrary rotation rate and other stellar parameters (such as molecular viscosity, particle diffusivity, etc).

In the solar context, many authors have investigated the structure of turbulence in rotating bodies to explain the occurrence of differential rotation in the convective zone. The main feature of this type of turbulence is the appearance of non-diffusive terms in the transport of angular momentum, which prevents a solid body rotation from being a  solution of the Reynolds equation \citep{Lebedinsky41,Kippenhahn63}. Starting from the Navier-Stokes equation, it is possible to show that these fluxes arise when there is a cause of anisotropy in the system, due to anisotropic background turbulence \citep[see][ and references therein]{Rudiger89} or inhomogeneities such as an underlying stratification \citep{Kichatinov87}. To explain the observed differential rotation of the sun, it is also of prime importance to understand the influence of rotation on the turbulent transport coefficients such as the turbulent diffusivity of particles and the turbulent conductivity of heat. \citet{Kichatinov94} have shown that for an isotropic original turbulence with long correlation time, all these coefficients scale as $\Omega^{-1}$ (where $\Omega$ is the rotation rate of the sun) in the large rotation limit. 

In our recent publications, we have studied the dynamics of the Sun by taking into account the presence of a strong shear \citep{Kim05,2Shears} and the interaction of this sheared turbulence with different types of waves that can be excited in the Sun due to magnetic fields \citep{BetaPlane,Dynamics}, stratification \citep{Stratification} or global rotation \citep{RotShearAA}. The main purpose of these studies was to understand the dynamics of the thin radial shear layer in the Sun, namely the solar tachocline \citep{Spiegel92}, where a sharp transition between latitudinal differential rotation in the convective envelope and nearly uniform rotation in the radiative interior takes place. 

In this paper, we consider the opposite limit where the differential rotation is negligible compared to the global rotation. This is relevant for the convection zone and also for rapidly rotating stars which are more massive than the Sun \citep{Kuker05}. Specifically, we investigate the effect of rotation on the properties of turbulence; i.e. its intensity and its transport. First, using a quasi-linear theory of turbulence with an isotropic arbitrary forcing, we show that all the turbulent transport coefficients, both parallel and perpendicular to the rotation vector, have the asymptotic scaling $\Omega^{-1}$ when the rotation rate is sufficiently large. The turbulence amplitude is however found to scale as $\Omega^0-\Omega^{-1}$ in the rapid rotation limit depending on the property of the forcing. We also consider the alternative case of an anisotropic turbulence with inhibited motion in the vertical direction, as should be relevant, for example, in a strongly stratified medium. In this case, we find that the Reynolds stress does not vanish (unlike in the isotropic case where it vanishes) and that the transport of angular momentum scales as $\Omega^{-2}-\Omega^{-1}$ in the rapid rotation limit. Secondly, we perform a multi-scale renormalisation group analysis to improve on the quasi-linear theory. The result suggests that the transport scales as $\Omega^{-1/2}$, with a weaker dependence on $\Omega$ compared to that found by the quasi-linear theory.

The rest of the paper is organised as follows: in Sect. \ref{Model}, we solve the quasi-linear equations for the fluctuating velocity and density of particles in a rotating frame with an arbitrary external forcing. We then calculate the turbulent intensity and turbulent transport  in the case of a homogeneous forcing with arbitrary temporal correlations in Sect. \ref{Turbcoeff}. In Sect. \ref{SecIsotropic} and \ref{SecAnisotropic}, we present results in the case of an isotropic and anisotropic forcing,  respectively. Section \ref{RNGAnalysis} is devoted to the multi-scale renormalisation group analysis. Finally, we summarise and discuss our results in Sect. \ref{Discussion} and provide the implications for the turbulence in stellar interiors in Sect. \ref{Implications}.

\section{Model (governing equations)}
\label{Model}
Our starting point is the Navier-Stokes equation with a forcing term ${\bf f}$ combined with an advection-diffusion equation for the transport of chemical species, in a rotating frame with angular velocity $\tilde{\bf \Omega}$:
\EQA
\label{NSrotation}
\partial_t \vv + \vv \cdot \nabla \vv &=& - \nabla p + \nu \nabla^2 \vv + {\bf f} - 2 \tilde{\bf \Omega} \times \vv \; , \\ \nonumber
\nabla \cdot \vv &=& 0 \; , \\ \nonumber
\partial_t N + \vv \cdot \nabla N &=& D \nabla^2 N \; ,  
\ENA
where $\nu$ is the viscosity of the fluid and $D$ the diffusivity of particle. Note that, in the incompressible case considered here, the heat transport can be described by an advection-diffusion equation \citep[e.g. see][]{Chandrasekhar}. Thus, our result for the turbulent diffusivity also holds for the turbulent conductivity, provided that $D$ is replaced by the thermal conductivity. In the following, we  let $ \OOmega = 2 \tilde{\bf \Omega}$ and assume (without loss of generality) that the rotation is around the $x$-axis with no large-scale velocity field \citep[note that the combined effect of rotation and large-scale shear is investigated in][]{RotShearAA}. We then express the velocity field and the concentration as the sum of a large-scale field and small-scale fluctuations: $\vv = \UU_0 + \uu = \uu$ and $N= N_0 + n$. In the quasi-linear theory \citep{Moffatt78}, Eq. (\ref{NSrotation}) can be linearised for the fluctuating fields and then Fourier-transformed to yield equations for the Fourier component $\tilde{X}({\bf k},t)$. To consider arbitrary Prandtl number $P_r = \nu / D$, we introduce the following new variables $\hat{X} = \exp[\nu k^2 t] \tilde{X}$ and $\breve{X} = \exp[D k^2 t] \tilde{X}$ to absorb the viscosity and diffusivity terms, respectively. Eliminating the pressure variable, the equations for the velocity field and concentration can be written:
\EQA
\label{PureRot}
\partial^2_t \hat{u}_x &+&  \omega_0^2  \hat{u}_x = \frac{a}{\gamma+a^2} \Bigl[\partial_t \frac{\hat{h}_1(t)}{a} - \Omega \hat{h}_2(t) \Bigr] \; , \\ \nonumber
\partial_t \hat{u}_z &=& - \frac{\beta a}{\gamma} \partial_t\hat{u}_x  + \frac{\Omega a}{\gamma} \hat{u}_x + \frac{\hat{h}_2(t)}{\gamma} \; , \\ \nonumber
\hat{u}_y &=& - (a \hat{u}_x  + \beta \hat{u}_z) \; , \\ \nonumber
\partial_t \breve{n} &=& - \breve{\uu} \cdot \nabla N_0 \; ,
\ENA
where $a=k_x/k_y$, $\beta = k_z/k_y$, $\gamma=1+\beta^2$ and $\omega_0^2=\Omega^2 a^2/(\gamma+a^2)$; $\hat{h}_1(t) = \gamma \hat{f_x}(t) - a \hat{f_y}(t) -\beta a \hat{f_z}(t)$ and $\hat{h}_2(t) = - \beta \hat{f_y}(t) + \hat{f_z}(t)$. The homogeneous solution of the first equation can easily be found in terms of trigonometric functions. Using these solutions and the method of variation of constants, the solutions to the first three equations of Eq. (\ref{PureRot}) can then be derived:
\EQA
\label{Solution}
\nonumber
\hat{u}_x &=& \int_0^t dt' \Bigl\{\frac{\hat{h}_1(t')}{\gamma+a^2} \cos[\omega_0(t'-t)] + \frac{\theta \hat{h}_2(t')}{\sqrt{\gamma+a^2}} \sin[\omega_0(t'-t)] \Bigr \} \; ,  \\ 
\hat{u}_y &=& \int_0^t \, dt' \, \Bigl\{\frac{\hat{h}_1(t') a}{\gamma(\gamma+a^2)} \Bigl(- \cos[\omega_0(t'-t)] \\ \nonumber
&& + \frac{\theta \beta \sqrt{\gamma+a^2}}{a}\sin[\omega_0(t'-t)] \Bigr)  +  \frac{\hat{h}_2(t')a}{\gamma\sqrt{\gamma+a^2}} \times \\ \nonumber
&& \Bigl(- \theta  \sin[\omega_0(t'-t)] - \frac{\sqrt{\gamma+a^2} \beta}{a}\cos[\omega_0(t'-t)] \Bigr) \Bigr \} \; , \\ \nonumber
\hat{u}_z &=& - \int_0^t \, dt' \, \Bigl\{\frac{\hat{h}_1(t') a}{\gamma(\gamma+a^2)} \Bigl(\beta \cos[\omega_0(t'-t)]  \\ \nonumber
&& + \frac{\theta \sqrt{\gamma+a^2}}{a}\sin[\omega_0(t'-t)] \Bigr) +  \frac{\hat{h}_2(t')a}{\gamma\sqrt{\gamma+a^2}} \times \\ \nonumber
&& \Bigl(\theta \beta \sin[\omega_0(t'-t)] - \frac{\sqrt{\gamma+a^2}}{a}\cos[\omega_0(t'-t)] \Bigr) \Bigr \} \; , 
\ENA
where $\theta$ is the sign of $(\Omega a)$. 

In the following sections, we use Eq. (\ref{Solution})  to calculate both the turbulet amplitude and transport (of chemicals or heat). To this end, we prescribe the correlation function of the forcing  to be spatially homogeneous with a temporal correlation $C(\tau)$:
\begin{equation}
\label{Forcing}
\langle \hat{h}_i({\bf k_1},t_1) \hat{h}_j({\bf k_2},t_2) \rangle = (2\pi)^3 \delta({\bf k_1}+{\bf k_2}) \, C(\vert \, t_1-t_2 \vert) \, \phi_{ij}({\bf k_2}) \; ,
\end{equation}
for $i$ and $j=1$ or $2$. The functions $\phi_{ij}$ are the power spectra of the forcing. 

\section{Turbulent amplitude and transport}
\label{Turbcoeff}
\subsection{Turbulent amplitude}
\label{SectIntensity}
Using Eqs. (\ref{Solution}) and (\ref{Forcing}), it is straightforward to obtain the intensity of turbulence as:
\EQA
\label{TurbInt}
\langle u_x^2 \rangle  &=& \frac{1}{(2\pi)^3} \int d^3 k \Bigl\{A^x_{11} \frac{\phi_{11}({\bf k})}{(\gamma+a^2)^2}  \\ \nonumber
&& \hskip1cm + A^x_{12} \frac{2 \, \phi_{12}({\bf k})}{(\gamma+a^2)^{3/2}} + A^x_{22} \frac{\phi_{22}({\bf k})}{(\gamma+a^2)} \Bigr\} \; , \\ \nonumber
\langle u_y^2 \rangle  &=& \frac{1}{(2\pi)^3} \int d^3 k \Bigl\{A^y_{11} \frac{a^2 \phi_{11}({\bf k})}{\gamma^2 (\gamma+a^2)^2} \\ \nonumber
&& \hskip1cm + A^y_{12} \frac{2 a^2  \, \phi_{12}({\bf k})}{\gamma^2 (\gamma+a^2)^{3/2}} + A^y_{22} \frac{a^2 \phi_{22}({\bf k})}{\gamma^2 (\gamma+a^2)} \Bigr\} \; .
\ENA
Here, the functions $A^i_{jk}(k,\omega_0)$ are defined in Eq. (\ref{BullShit150}) of Appendix \ref{GrossesExpr}: they depend only on $k$ and $\Omega_0$ and involve the temporal correlation function of the forcing $C(\tau)$ only through the functions $\kappa_c$ and $\sigma_c$, defined by:
\EQA
\label{BullShit10}
\kappa_c &=& \int_0^{+\infty} e^{-\nu k^2 \tau}  \cos(\omega_0 \tau) \, C(\tau) \, d\tau \; ,\\ \nonumber
\sigma_c &=& \int_0^{+\infty} e^{-\nu k^2 \tau} \sin(\omega_0 \tau) \, C(\tau) \, d\tau \; .
\ENA
The turbulent amplitude in the $z$ direction $\langle u_z^2 \rangle$ can easily be obtained from $\langle u_y^2 \rangle$, due to the symmetry of the problem, by making the following replacements:
\EQA
\label{Symmetry}
\beta &\rightarrow& \beta^{-1} \; , \qquad a \rightarrow  a \, \beta^{-1} \; , \qquad \gamma \rightarrow \gamma \, \beta^{-2} \; , \\ \nonumber
\phi_{11} &\rightarrow& \phi_{11} \, \beta^{-4} \; , \qquad \phi_{12} \rightarrow \phi_{12} \, \beta^{-3} \; , \qquad \phi_{22} \rightarrow \phi_{22} \, \beta^{-2} \; .
\ENA

\subsection{Transport of angular momentum}
From Eqs. (\ref{Solution}) and (\ref{Forcing}), we can also calculate the non-diagonal part of the correlation tensors $\langle u_i u_j \rangle$ to obtain the transport of angular momentum by turbulence:
\EQA
\label{TranspAngular}
\langle u_x u_y \rangle  &=& \frac{1}{(2\pi)^3} \int d^3 k \Bigl\{M^x_{11} \frac{a \phi_{11}({\bf k})}{\gamma(\gamma+a^2)^2} \\ \nonumber
&& \hskip1cm + M^x_{12} \frac{2 a  \, \phi_{12}({\bf k})}{\gamma (\gamma+a^2)^{3/2}} + M^x_{22} \frac{a \phi_{22}({\bf k})}{\gamma (\gamma+a^2)} \Bigr\} \; , \\ \nonumber
\langle u_y u_z \rangle  &=& \frac{1}{(2\pi)^3} \int d^3 k \Bigl\{M^z_{11} \frac{a^2 \phi_{11}({\bf k})}{\gamma^2(\gamma+a^2)^2} \\ \nonumber
&& \hskip1cm + M^z_{12} \frac{2 a^2 \, \phi_{12}({\bf k})}{\gamma^2 (\gamma+a^2)^{3/2}} + M^z_{22} \frac{a^2 \phi_{22}({\bf k})}{\gamma^2 (\gamma+a^2)} \Bigr\} \; .
\ENA
Here, the functions $M^x_{ij}$ and $M^z_{ij}$ are given in Eq. (\ref{BullShit150}) of appendix \ref{GrossesExpr} and depend only on $k$, $\Omega_0$ and the functions $\kappa_c$ and $\sigma_c$ defined in Eq. (\ref{BullShit10}).

\subsection{Transport of particles}
Integrating the last equation of Eq. (\ref{Solution}), we obtain the density of chemical species:
\EQ
\label{Concentration}
\breve{n}_i = (-\partial_i N_0) \int_0^t  \, \breve{u}_i(t') \, dt' \; .  
\EN
Note that the velocity to be integrated in Eq. (\ref{Concentration}) differs by an exponential factor from that given by Eq. (\ref{Solution}). However it is more convenient to express the turbulent diffusivity of chemical species $D_T^{ij}$, defined by  $\langle u_i n \rangle = -  D_T^{ij} \partial_j N_0$, in terms of $\hat{u}_i$. Specifically, we can obtain the following:
\EQA
\label{TranspVert}
D_T^{ij} &=& \frac{1}{(2 \pi)^6} \int d^3 k_1 \int d^3 k_2 e^{-(D k_1^2 + \nu k_2^2)t} \times \\ \nonumber 
&& \int_0^t \, dt' \,  e^{(D-\nu) k_1^2 t'} \langle \hat{u}_j({\bf k_1},t') \hat{u}_i({\bf k_2},t) \rangle \; .
\ENA
Using Eqs. (\ref{Solution})-(\ref{Forcing}) and following a calculation similar to that in Sect. \ref{SectIntensity}, we can calculate all the components of the turbulent diffusivity $D_T^{ij}$. The results turn out to depend on the temporal correlation function $C(\tau)$ through a new function:
\EQ
\label{BullShit11}
\zeta_c = \int_0^{+\infty} e^{- D k^2 \tau} C(\tau) \, d\tau \; .
\EN

These results in a general form are not shown here for the sake of brevity and shall be presented in the case of isotropic and anisotropic forcing in Sect. \ref{SecIsotropic} and  \ref{SecAnisotropic}, respectively.

\section{Isotropic forcing}
\label{SecIsotropic}
When the forcing ${\bf f}$ is isotropic, the correlation function of the forcing (with no helicity) can be written as: 
\EQ
\label{IsotropicFor}
\langle f_i({\bf k},t) f_j({\bf k'},t') \rangle = (2 \pi)^3 \, C(\vert t-t' \vert) \, F(k) \, (\delta_{ij} - k_i k_j / k^2 ) \, \delta({\bf k}+{\bf k'}) \; .
\EN 
The functions $\phi_{ij}$ in Eq. (\ref{Forcing}) are related to $F(k)$ in (\ref{IsotropicFor}) by:
\EQ
\label{BullShit01}
\phi_{11} = \gamma (\gamma+a^2) F(k) \quad , \quad \phi_{12} = 0 \quad \mathrm{and} \quad \phi_{22} = \gamma F(k) \; .
\EN
Using Eq. (\ref{BullShit01}) in Eqs. (\ref{TurbInt}), (\ref{TranspAngular}) and (\ref{TranspVert}), we can obtain the turbulence amplitude and transport as follows:
\EQA
\langle u_x^2 \rangle  &=& \frac{1}{(2\pi)^3} \int d^3 k  \frac{\gamma F(k)}{(\gamma+a^2)} \frac{\kappa_c}{\nu k^2} \; , \\ \nonumber
\langle u_y^2 \rangle  &=& \frac{1}{(2\pi)^3} \int d^3 k  \frac{a^2 F(k)}{\gamma (\gamma+a^2)} \left\{ \frac{\gamma(\beta^2+a^2)}{\nu k^2 a^2} \kappa_c  +  \frac{\omega_0}{\nu^2 k^4 + \omega_0^2} \sigma_c \right\}  \; , \\ \nonumber 
\langle u_x u_y \rangle &=& - \frac{1}{(2\pi)^3} \int d^3 k  \frac{a F(k)}{(\gamma+a^2)} \frac{\kappa_c}{\nu k^2}  \; , \\ \nonumber
\langle u_y u_z \rangle &=& - \frac{1}{(2\pi)^3} \int d^3 k  \frac{F(k)}{(\gamma+a^2)} \frac{\beta}{\nu k^2} \kappa_c \; , \\ \nonumber
D_T^{xx} &=& \frac{1}{(2\pi)^3} \int d^3 k \frac{\gamma F(k)}{(\gamma+a^2)} \mathcal{I}(k,\omega_0) \; , \\ \nonumber
D_T^{yy} &=& \frac{1}{(2\pi)^3} \int d^3 k \frac{F(k) (\beta^2+a^2)}{(\gamma+a^2)}  \mathcal{I}(k,\omega_0) \; .
\ENA
Here, the functions $\chi_c$, $\kappa_c$ and $\sigma_c$ [defined in Eqs. (\ref{BullShit10}) and (\ref{BullShit11})] depend only on the modulus of the wave number; The function $\mathcal{I}(k,\omega_0)$ is defined as:
\EQA
\mathcal{I}(k,\omega_0) &=& \frac{1}{(D-\nu)^2 k^4 + \omega_0^2} \Bigl[\frac{D-\nu}{\nu} \kappa_c \\ \nonumber
&& - \frac{(D^2-\nu^2) k^4 - \omega_0^2}{(D + \nu)^2 k^4 + \omega_0^2}  (\kappa_c+\zeta_c) + \frac{2 D k^2 \omega_0}{(D + \nu)^2 k^4 + \omega_0^2} \sigma_c \Bigr]  \; .
\ENA
By performing the integration in the $y$-direction and changing the colatitude variable to $\omega_0 = \vert \Omega \cos\theta \vert$, we obtain the following result:
\EQA
\label{Isotropic}
\nonumber
\langle u_x^2 \rangle  &=& \frac{2}{(2\pi)^2 \vert \Omega \vert} \int_0^{+\infty} dk \; k^2 F(k) \int_{0}^{\vert \Omega \vert} d\omega_0 \left(1-\frac{\omega_0^2}{\Omega^2}\right) \frac{\kappa_c(k,\omega_0)}{\nu k^2} \; , \\ 
\langle u_y^2 \rangle  &=&\frac{1}{(2\pi)^2 \vert \Omega \vert} \int_0^{+\infty} dk \; k^2 F(k) \int_{0}^{\vert \Omega \vert} d\omega_0  \times \\ \nonumber
&& \left\{ \left(1+\frac{\omega_0^2}{\Omega^2}\right) \frac{\kappa_c(k,\omega_0)}{\nu k^2}   +  \frac{\omega_0^2}{\Omega^2} \frac{\omega_0}{\nu^2 k^4 + \omega_0^2} \sigma_c(k,\omega_0) \right\}  \; , \\ \nonumber 
D_T^{xx} &=& \frac{2}{(2\pi)^2 \vert \Omega \vert} \int_0^{+\infty} dk \; k^2 F(k) \int_{0}^{\vert \Omega \vert} d\omega_0 \left(1-\frac{\omega_0^2}{\Omega^2}\right) \mathcal{I}(k,\omega_0) \; , \\ \nonumber
D_T^{yy} &=& \frac{1}{(2\pi)^2 \vert \Omega \vert} \int_0^{+\infty} dk \; k^2 F(k) \int_{0}^{\vert \Omega \vert} d\omega_0 \left(1+\frac{\omega_0^2}{\Omega^2}\right) \mathcal{I}(k,\omega_0) \; ,
\ENA
and vanishing transport of angular momentum: $\langle u_x u_y \rangle = \langle u_x u_y \rangle = 0$. We now examine the dependence of the turbulence amplitude and transport of particles in Eq. (\ref{Isotropic}) on the correlation time of the forcing $C(\tau)$.

\subsection{Infinitely correlated turbulence}
\label{IsotropInfinite}
When $C(\tau)=1$, the correlation functions in Eqs. (\ref{BullShit10}) and (\ref{BullShit11}) take the following forms:
\EQ
\label{Infinicorrele}
\zeta_c = \frac{1}{D k^2} \quad , \quad \kappa_c = \frac{\nu k^2}{\nu^2 k^4 + \omega_0^2} \quad \mathrm{and} \quad \sigma_c = \frac{\omega_0}{\nu^2 k^4 + \omega_0^2} \; .
\EN
 Using Eq. (\ref{Infinicorrele}) in Eq. (\ref{Isotropic}) and performing the integration on the $\omega_0$ variable, we obtain the following equations:
\EQA
\label{Rudiger}
\nonumber
\langle u_x^2 \rangle  &=& \frac{2}{(2\pi)^2} \int_0^{+\infty} dk \; \frac{F(k)}{\nu^2 k^2}  \frac{1}{\Omega_*^2} \left(-1+\frac{\Omega_*^2+1}{\Omega_*}\arctan\Omega_*\right) \; , \\ 
\langle u_y^2 \rangle  &=& \frac{1}{(2\pi)^2} \int_0^{+\infty} dk \; \frac{F(k)}{\nu^2 k^2}  \frac{1}{\Omega_*^2} \times \\ \nonumber
&& \left(2+\frac{1}{2(1+\Omega_*^2)}+\frac{2\Omega_*^2-5}{2\Omega_*}\arctan\Omega_* \right)   \; , \\ \nonumber 
D_T^{xx} &=& \frac{2}{(2\pi)^2} \int_0^{+\infty} dk \; \frac{F(k)}{D \nu^2 k^4}  \frac{1}{\Omega_*^2} \left(-1+\frac{\Omega_*^2+1}{\Omega_*}\arctan\Omega_*\right)  \; , \\ \nonumber
D_T^{yy} &=& \frac{1}{(2\pi)^2} \int_0^{+\infty} dk \; \frac{F(k)}{D \nu^2 k^4}  \frac{1}{\Omega_*^2} \left(1+\frac{\Omega_*^2-1}{\Omega_*}\arctan\Omega_*\right)  \; ,
\ENA
where $\Omega_* = \Omega_*(k) = \vert \Omega \vert / (\nu k^2)$. In the limit $\Omega_* \rightarrow \infty$, the turbulent transport and intensity, both for the parallel and perpendicular components, all tend to zero as $\Omega^{-1}$. These agree with \citet{Kichatinov94}. The effect of rotation on turbulence amplitude appears with $\Omega_*$, which becomes large on large scales. In other words, turbulence quenching is more severe on large scales, leading to effectively strong turbulence on small scales. This is consistent with the reduction of scale of motion as evidenced in numerical simulations \citep{Brummell98}.

Interestingly, the turbulent transport of particles is proportional to $D^{-1}$, suggesting that the turbulent transport of heat scales as $\kappa^{-1}$ with the molecular heat diffusivity $\kappa$. Thus, in the Sun where $\kappa \gg D$, the transport of particles is expected to be faster than that of heat by a factor $\kappa / D \sim 10^7 / 10^2 \sim 10^5$. However, this holds only for an incompressible fluid, as the equation for the transport of heat is not the same as that of particles for a compressible fluid \citep{Spiegel60}. The relevance of this result for the solar context may thus be questionnable.

\subsection{$\delta$-correlated turbulence}
\label{IsotropDelta}
We now consider a short correlated turbulence modelled by $C(\tau) = \tau_f \delta(\tau)$. In this case, the temporal correlations in Eqs. (\ref{BullShit10}) and (\ref{BullShit11}) become:
\EQ
\label{Deltacorrele}
\zeta_c = \kappa_c =\frac{\tau_f}{2} \quad \mathrm{and} \quad \sigma_c = 0 \; ,
\EN
leading to the following turbulent intensity and transport:
\EQA
\label{Isotropic2}
\langle u_x^2 \rangle &=& \langle u_y^2 \rangle = \frac{2}{3(2 \pi)^2} \int_0^{\infty} \frac{F(k) \tau_f}{\nu} \; dk \; , \\ \nonumber
D_T^{xx} &=& \frac{1}{(2 \pi)^2} \int_0^{\infty} \frac{F(k) \tau_f}{\nu^2 k^2} \frac{1}{\Omega_*^2} \times \\ \nonumber
&& \left[-(1+b)+\frac{\Omega_*^2+(1+b)^2}{2 \Omega_*}\arctan\left(\frac{\Omega_*}{2}\right) \right] \; dk \; , \\ \nonumber
D_T^{yy} &=& \frac{1}{2(2 \pi)^2} \int_0^{\infty} \frac{F(k) \tau_f}{\nu^2 k^2} \frac{1}{\Omega_*^2} \times \\ \nonumber
&& \left[1+b+\frac{\Omega_*^2-(1+b)^2}{2 \Omega_*}\arctan\left(\frac{\Omega_*}{2}\right) \right] \; dk \; .
\ENA
Here, $b=1/Pr$. Eq. (\ref{Isotropic2}) shows that, in the case of a short correlated forcing, only the turbulent transport is suppressed (by a factor $\Omega^{-1}$) whereas the turbulent intensity is of the same order as that without rotation. This is because, in this case, inertial waves only have an effect on the phase of the velocity field and thus do not modify its amplitude. Alternatively, this is because the flow driven by a forcing with a short correlation time  $\tau_f < \Omega^{-1}$ has no coherent motion to be affected by inertial waves (or rotation). This is to be compared with the result obtained in the case of an infinitely correlated forcing (Sect. \ref{IsotropInfinite}) where the turbulence amplitude was reduced as $\Omega^{-1}$ due to rotation. Furthermore, the dependence of the turbulent diffusivity on the molecular diffusivity is not as simple as in the infinitely correlated case. However, in the large rotation limit ($\Omega_* \gg 1$), we can easily see from Eq. (\ref{Isotropic2}) that the turbulent diffusivity depends neither on $P_r$ nor on $D$. Consequently, the transport of heat and particle is the same for rapid rotation.

\subsection{Finite correlation time with exponential correlation function}
\label{Isotropicfinitecorrelation}
In Sect. \ref{IsotropInfinite} and \ref{IsotropDelta}, we considered two extreme limits where the correlation time was infinite or zero, respectively. Here, we discuss a more realistic case where the turbulence has a finite correlation time $\tau_f$ by using an exponential correlation function $C(\tau) = \exp[-2 \tau / \tau_f]$. In this case, the functions in Eqs. (\ref{BullShit10}) and (\ref{BullShit11}) can be simplified as:
\EQA
\label{BullShit20}
\label{Correlation Exp}
\kappa_c &=& \frac{\nu k^2+2/\tau_f}{(\nu k^2 + 2/\tau_f)^2+\omega_0^2} \; , \\ \nonumber
\sigma_c &=& \frac{\omega_0}{(\nu k^2 + 2/\tau_f)^2+\omega_0^2} \; , \\ \nonumber
\zeta_c &=& \frac{1}{D k^2 + 2/\tau_f} \; .
\ENA
Putting Eq. (\ref{BullShit20}) in Eq. (\ref{Isotropic}), we can calculate the turbulent intensity and transport as previously. We omit, however, the general form of the results which are too complex, and only provide a result that is valid in the limit $\Omega \rightarrow \infty$.  To this end, we can simply take the limit $\Omega \rightarrow \infty$ in Eq. (\ref{Isotropic}) and obtain the following result:
\EQA
\label{BullShit21}
\langle u_x^2 \rangle  &\sim& \frac{2}{(2\pi)^2 \vert \Omega \vert} \int_0^{+\infty} dk \; k^2 F(k) \int_{0}^{\infty} d\omega_0 \frac{\kappa_c(k,\omega_0)}{\nu k^2} \; , \\ \nonumber
\langle u_y^2 \rangle  &\sim& \frac{1}{(2\pi)^2 \vert \Omega \vert} \int_0^{+\infty} dk \; k^2 F(k) \int_{0}^{\infty} d\omega_0 \frac{\kappa_c(k,\omega_0)}{\nu k^2}  \; , \\ \nonumber 
D_T^{xx} &\sim& \frac{2}{(2\pi)^2 \vert \Omega \vert} \int_0^{+\infty} dk \; k^2 F(k) \int_{0}^{\infty} d\omega_0 \, \mathcal{I}(k,\omega_0) \; , \\ \nonumber
D_T^{yy} &\sim& \frac{1}{(2\pi)^2 \vert \Omega \vert} \int_0^{+\infty} dk \; k^2 F(k) \int_{0}^{\infty} d\omega_0  \, \mathcal{I}(k,\omega_0)  \; .
\ENA
We note that, in order to obtain Eq. (\ref{BullShit21}), the limit $\Omega \rightarrow \infty$ was taken inside the integral as the resulting integrals are all convergent. Eq. (\ref{BullShit21}) shows that both turbulent amplitude and transport scale  as $\Omega^{-1}$ for large rotation rate. This is similar to the case of the infinite correlation time case. Furthermore, Eq. (\ref{BullShit21}) shows that the ratio between the parallel and perpendicular transport of particles (or heat) is given by:
\EQ
\label{Relation}
\frac{\langle u_x^2 \rangle}{\langle u_y^2 \rangle} = \frac{D_T^{xx}}{D_T^{yy}} = 2 \quad \mathrm{for} \quad \Omega \rightarrow \infty \; ,
\EN
again similar to the case of $C(\tau) =1$. We note that the large rotation limit $\Omega \rightarrow \infty$ taken above requires that not only $\Omega \gg \nu k^2$ but also $\Omega \gg \tau_f^{-1}$. In other words, the correlation time of the forcing should be larger than $\Omega^{-1}$. In the solar context, $\Omega \sim 3\times10^{-6} s$. Thus $\tau_f$ must be larger than a few days. When $\Omega < \tau_f^{-1}$, it can easily be shown that the results become similar to those in the case of $\delta$-correlated forcing studied in Sect. \ref{IsotropDelta} (i.e. $\langle u^2 \rangle \sim \Omega^{0}$ and $D_T \sim \Omega^{-1}$).

\subsection{Power law correlation}
\label{PowerlawSection}
In view of the highly intermittent nature of stellar activity, it is interesting to consider the case with a correlation function of the forcing as a power-law. For simplicity, we consider a correlation function given by:
\EQ
C(\tau) = \left(\frac{\tau_f}{\tau}\right)^{\mu} \; ,
\EN
with $0 < \mu < 1$ (note that it is also possible to consider a function with exponent $\mu > 1$ by introducing a cutoff frequency). To calculate the various correlation functions, we need the following integral:
\EQ
I = \int_0^{+\infty} e^{(i \omega_0 - \nu k^2) \tau} \tau^{-\mu} d\tau \; = \left(\frac{i}{\nu k^2}\right)^{1-\mu} \int_0^{+\infty} t^{-\mu} e^{-it-\omega_0^* t} \; dt \; , 
\EN
where the last equation is obtained using steepest descent method. Using Watson's lemma \citep{Bender}, the asymptotic behaviour of this integral for $\omega_0^* = \omega_0 / \nu k^2 \gg 1$ can be found as:
\EQ
I \sim \Gamma(1-\mu) \left(\frac{i}{\omega_0}\right)^{1-\mu} \; .
\EN
Thus the correlation functions in Eqs. (\ref{BullShit10}) and (\ref{BullShit11}) become:
\EQA
\label{Powerlaw}
\zeta_c &=& \frac{\Gamma(1-\mu)}{D k^2} \tau_{f*}^\mu \, , \\ \nonumber
\kappa_c &\sim& \frac{\Gamma(1-\mu)  \tau_{f*}^\mu}{\nu k^2} \omega_{0*}^{\mu-1} \cos\left[(1-\mu) \frac{\pi}{2}\right] \; , \\ \nonumber
\sigma_c &\sim& \frac{\Gamma(1-\mu)  \tau_{f*}^\mu}{\nu k^2} \omega_{0*}^{\mu-1} \sin\left[(1-\mu) \frac{\pi}{2}\right]  \; ,
\ENA 
where $\tau_{f*}=\tau_f \nu k^2$ and $\omega_{0*}=\omega_{0} / \nu k^2$ are the correlation time and the oscillation frequency of the noise scaled by the diffusion time, respectively. Putting Eq. (\ref{Powerlaw}) in Eq. (\ref{Isotropic}), we then obtain the turbulence amplitude and turbulent transport of chemicals for an isotropic forcing with a power-law correlation:
\EQA
\label{ResultPower}
\langle u_x^2 \rangle  &=& \frac{2}{(2\pi)^2} \int_0^{+\infty} dk \; \frac{F(k) \tau_{f*}^\mu}{\nu^2 k^2 \Omega_*} \cos\left[(1-\mu) \frac{\pi}{2}\right] \frac{2 \Omega_*^\mu}{\mu(\mu+2)}  \; , \\ \nonumber
\langle u_y^2 \rangle  &=&\frac{1}{(2\pi)^2 } \int_0^{+\infty} dk \; \frac{F(k) \tau_{f*}^\mu}{\nu^2 k^2 \Omega_*} \cos\left[(1-\mu) \frac{\pi}{2}\right] \frac{2 (\mu+1) \Omega_*^\mu}{\mu(\mu+2)} \; , \\ \nonumber 
D_T^{xx} &=& \frac{2}{(2\pi)^2 } \int_0^{+\infty} dk \; \frac{F(k) \tau_{f*}^\mu}{D \nu^2 k^4 \Omega_*} T\; , \\ \nonumber
D_T^{yy} &=& \frac{1}{(2\pi)^2 } \int_0^{+\infty} dk \; \frac{F(k) \tau_{f*}^\mu}{D \nu^2 k^4 \Omega_*} T \; ,
\ENA
in the rapid rotation limit $\Omega_* = \vert \Omega \vert / (\nu k^2) \gg 1$. $T$ is a constant which is independent of $\Omega$. From Eq. (\ref{ResultPower}), we can see that the turbulence intensity scales as $\vert \Omega \vert^{\mu-1}$ for rapid rotation. This shows that the scaling of the turbulent intensity depends sensitively on the exponent of the power law. Note that the result of infinitely correlated turbulence is recovered for $\mu = 0$. In comparison, the turbulent diffusivity in both directions scales as $\vert \Omega \vert^{-1}$ independent of the power exponent $\mu$. Furthermore, we can see that the ratio between the turbulence in the $x$- and $y$-direction depends only on the power exponent:
\EQ
\frac{\langle u_x^2 \rangle}{\langle u_y^2 \rangle} \sim \frac{2}{\mu+1} \; .
\EN
Consequently, the turbulence is always more vigorous in the $x$-direction (along the rotation axis) than in the $y$-direction. We recover the factor 2 for $\mu = 0$, corresponding to the exponentially correlated case, see Eq. (\ref{Relation}). As the power exponent is increased, the turbulence becomes more isotropic. Note that the ratio between the transport of species in the $x$- and $y$-direction is always 2, as in the case of an exponential correlation function.

To summarise, for an isotropic forcing and rapid rotation, we showed that the turbulent diffusivity is reduced as $\Omega^{-1}$, irrespective of the correlation time of the forcing,  with the transport in the direction parallel to the rotation vector being twice as fast as than in the direction perpendicular to the rotation. This would lead to a warmer pole than equator \citep{Kichatinov95}. In comparison, the turbulent intensity is found to be reduced with different scaling with $\Omega$ depending on the properties of the correlation function of the forcing. Specifically,
in the case of a finite correlation time, the turbulence intensity scales as  $\Omega^{-1}$ for rapid rotation, while it is independent of the rotation rate for a short-correlated forcing (Sect. \ref{IsotropDelta}). Furthermore, for a temporal correlation function given by a power-law, turbulence amplitude scales as $\Omega^n$ with $-1 < n < 0$.

\section{Anisotropic forcing}
\label{SecAnisotropic}
We now consider the case where the forcing ${\bf f}$ is anisotropic as should be relevant for the turbulence in the convection zone (for instance, due to the underlying stratification). We are particularly interested in the Reynolds stress as the combination of anisotropy and rotation can lead to non-diffusive fluxes of angular momentum. In the case where ${\bf g}$ is a unit vector in the direction of anisotropy, the correlation function of the forcing can be written \citep{Rudiger89}: 
\EQA
\label{Anisotropic}
\langle f_i({\bf k},t) f_j({\bf k'},t') \rangle = (2 \pi)^3 \, C(\vert t-t' \vert) \, \delta({\bf k}+{\bf k'}) \, G(k) \times \\ \nonumber
 \left[\delta_{ij} - \frac{k_i k_k}{k^2} - \frac{({\bf g} \cdot {\bf k})^2}{k^2} \delta_{ij} -  g_i g_j + \frac{{\bf g} \cdot {\bf k}}{k^2} (g_i k_j + g_j k_i) \right] \; .
\ENA 
This forcing in Eq. (\ref{Anisotropic}) is highly anisotropic with only a non-vanishing component perpendicular to ${\bf g}$:
\EQA
\langle ({\bf f} \cdot {\bf g})^2 \rangle &=& 0 \; , \\ \nonumber
\langle ({\bf f} \times {\bf g})^2 \rangle &=& \frac{1}{2 \pi^2} \int_0^{\infty} G(k) k^2 dk \; .
\ENA
In the following, we assume ${\bf g}$ to be in the vertical direction. In general, one can consider a less anisotropic forcing by combining the isotopic part of the forcing (\ref{IsotropicFor}) and anisotropic part (\ref{Anisotropic}).

\subsection{Rotation parallel to the vertical direction}
When the rotation vector is parallel to the vertical direction i.e. ${\bf g} \parallel {\bf \Omega}$ (as in the case of the turbulence near the pole), Eq. (\ref{Anisotropic}) translates to:
\EQ
\label{BullShit30}
\phi_{11} = \phi_{12} = 0 \quad \mathrm{and} \quad \phi_{22} = \frac{\gamma^2}{\gamma+a^2} G(k) \; .
\EN
Using Eq. (\ref{BullShit30}) in Eqs. (\ref{TurbInt}) and (\ref{TranspAngular}), we can show that the turbulence amplitude becomes
\EQA
\nonumber
\langle u_x^2 \rangle  &=& \frac{2}{(2\pi)^2 \vert \Omega \vert} \int_0^{\infty} d k \, k^2 G(k) \int_0^{\vert \Omega \vert} d \omega_0 \left(1-\frac{\omega_0^2}{\Omega^2} \right)^2 \times \\ \nonumber
&& \left[\left( \frac{1}{2\nu k^2} - \frac{\nu k^2}{2(\nu^2 k^4 + \omega_0^2)} \right) \kappa_c + \frac{\omega_0}{2(\nu^2 k^4 + \omega_0^2)} \sigma_c \right]  \; , \\ \label{BullShit31}
\langle u_y^2 \rangle  &=& \frac{1}{(2\pi)^2 \vert \Omega \vert} \int_0^{\infty} d k \, k^2 G(k) \int_0^{\vert \Omega \vert} d \omega_0  \left(1-\frac{\omega_0^2}{\Omega^2} \right) \times \\ \nonumber
&& \left[\left(1+\frac{\omega_0^2}{\Omega^2} \right) \frac{\kappa_c}{2\nu k^2} + \frac{\nu k^2 \kappa_c - \omega_0 \sigma_c}{2(\nu^2 k^4 + \omega_0^2)} \right] \; , 
\ENA
while the transport of angular momentum vanishes. Here again, we can see by using Eq. (\ref{Correlation Exp}) that all the integrals in Eq. (\ref{BullShit31}) converge in the limit $\Omega \rightarrow \infty$. Thus, turbulence amplitudes are reduced as $\Omega^{-1}$ for large rotation rate. We do not provide here the results for the turbulent diffusivities as they are very similar to those derived in the isotopic case. Thus, in the rapid rotation limit, turbulent diffusivities scale as $\Omega^{-1}$ for all the different types of correlation function of the forcing that were considered previously.

\subsection{Rotation perpendicular to the vertical direction}
Near the equator, the rotation vector is perpendicular to the vertical direction. Since the rotation is assumed to be in the $x$-direction, the vertical direction for anisotropy can be chosen to be in the $z$-direction without loss of generality. Eq. (\ref{Anisotropic}) then translates to:
\EQA
\label{BullShit40}
\phi_{11} = (\gamma+a^2) G(k) \quad , \quad \phi_{12} = a \beta G(k) \\ \nonumber 
\quad \mathrm{and} \quad \phi_{22} = \frac{a^2 \beta^2}{\gamma+a^2} G(k) \; .
\ENA

In this case, the turbulent amplitude and diffusivities can be shown to be very similar to those derived in the isotopic case. Consequently, the scalings are the same as those derived in Sect. \ref{SecIsotropic}. However, in this case, the transport of angular momentum does not vanish as discussed below. Putting Eq. (\ref{BullShit40}) into Eq. (\ref{TranspAngular}) and keeping only the even terms in $\beta$ and $a$ (as the odd terms vanish after angular integration), straightforward but cumbersome calculations lead to the following expressions for the  correlation function of the $y$ and $z$ components of the velocity:
\EQA
\label{BullShit160}
\langle u_y u_z \rangle = - \frac{1}{2 (2 \pi)^3} \int d^3 k \frac{\theta a (1+a^2) G(k)}{(\gamma+a^2)^{3/2}} \frac{\omega_0 \kappa_c + \nu k^2 \sigma_c}{\nu^2 k^4+ \omega_0^2} \; ,
\ENA
while the other correlation functions vanish. Performing the integration over the angular variables, Eq. (\ref{BullShit160}) can be simplified as:
\EQA
\label{BullShit170}
\langle u_y u_z \rangle &=& - \frac{\chi}{2(2 \pi)^2 \Omega^2} \int_0^{\infty} d k \, k^2 G(k)  \int_0^{\vert \Omega \vert} d \omega_0 \; \times \\ \nonumber
&& \omega_0 \left(1+\frac{\omega_0^2}{\Omega^2}\right) \frac{\omega_0 \kappa_c + \nu k^2 \sigma_c}{\nu^2 k^4+ \omega_0^2} \; ,
\ENA
where $\chi$ is the sign of $\Omega$. The Reynolds stress in Eq. (\ref{BullShit160}) depends on the direction of the rotation, representing non-diffusive momentum transport, i.e., $\Lambda$ effect \citep{Rudiger89}. This effect favours the creation of velocity gradient rather than smoothing it out and can thus be responsible for the differential rotation of stars. 

We now examine the behaviour of Eq. (\ref{BullShit160}) for different temporal correlation functions. In the case of an infinitely correlated forcing, Eq. (\ref{BullShit170}) can be shown to become:
\EQA
\label{BullShit50}
\langle u_y u_z \rangle &=& - \frac{\chi}{2 (2 \pi)^2} \int_0^{\infty} d k \, \frac{G(k)}{\nu^2 k^2  \Omega_*^3}  \times \\ \nonumber
&& \left[\frac{3+\Omega_*^2}{1+\Omega_*^2} + \frac{\Omega_*^2-3}{\Omega_*} \arctan{\Omega_*} \right] \; .
\ENA
Eq. (\ref{BullShit50}) shows that the non-diffusive fluxes of angular momentum scale as $\Omega^{-2}$ in the large rotation limit. This agrees with \citet{Kichatinov86b}.  Similarly, we can show that for a finite correlation time (with exponential form), the limit $\Omega_* \rightarrow \infty$ can be taken inside the integral of Eq. (\ref{BullShit170}) and consequently, the non-diffusive fluxes of angular momentum scale also as $\Omega^{-2}$ in the large rotation limit.

In comparison, when the forcing is $\delta$-correlated, Eq. (\ref{BullShit170})  becomes:
\EQA
\label{BullShit51}
\langle u_y u_z \rangle &=& - \frac{\chi}{2 (2 \pi)^2} \int_0^{\infty} d k \, \frac{\tau_f G(k)}{6 \nu \Omega_*^4}  \times \\ \nonumber
&& \left[\Omega_* (4 \Omega_*^2 - 3) + 3 (1-\Omega_*^2) \arctan{\Omega_*} \right] \; ,
\ENA
which shows that the non-diffusive fluxes of angular momentum scale as $\Omega^{-1}$ in the rapid rotation limit. 

Finally, for a power-law correlation function, by putting Eq. (\ref{Powerlaw}) into Eq. (\ref{BullShit170}), we obtain the non-diffusive fluxes of angular momentum as:
\EQA
\langle u_y u_z \rangle &=& - \frac{\chi}{2 (2 \pi)^2} \int_0^{\infty} d k \, \frac{\tau_{f*}^\mu G(k)}{\nu k^2 \Omega_*^2}  \Gamma(1-\mu)  \times \\ \nonumber
&& \frac{\Omega_*^\mu \cos\left[(1-\mu) \pi / 2 \right]}{2(\mu+2) \cos\left[(1+\mu) \pi / 2 \right]}  \left( \cos\left[(1+\mu) \frac{\pi}{2}\right] - \frac{1}{\mu} \right)\; ,
\ENA
in the rapid rotation limit. In this limit, the $\Lambda$-effect scales as $\Omega^{\mu-2}$. Thus,  for a power-law correlation function, the $\Lambda$-effect scales as $\Omega^{-n}$ with $1 < n <2$.

In summary, we show that non-diffusive fluxes of angular momentum do not vanish only when the rotation is perpendicular to the inhomogeneity (such as stratification in the radial direction or shear in the azimuthal direction). This is in agreement with previous results. Furthermore, we show that these non-diffusive fluxes scale as $\Omega^{-2}-\Omega^{-1}$, in the rapid rotation limit, depending on the property of the temporal correlation function of the forcing.

\section{Multi-scale renormalisation group analysis}
\label{RNGAnalysis}
The results presented in the previous sections are derived using the quasi-linear theory. This is strictly valid only for two-scale turbulence, with a spatial gap between a large-scale for the mean-flow and a small scale for the fluctuations, and for weak turbulence. Therefore, the applicability of this approximation is limited. In this section, we utilise a renormalisation group technique \citep[see][ and references therein for details]{Moffatt81} to consider a more realistic situation of a strong turbulence with a no-clear scale separation between turbulence and mean fields. This method is based on the computation of the turbulent diffusivity iteratively by considering a range of scales $l_n's$ with:
\EQ
l_n > l_{n-1} > \dots > l_0 \; , 
\EN
and $l_0$ is the scale at which the molecular diffusion becomes dominant. The effective turbulent diffusivity $D_{n+1}$ for scale $n+1$ is then calculated by using the value of the turbulent diffusivity at the preceding step, i.e., $D_n$ via quasi-linear theory.  Since the turbulent diffusivity scales as $D^{-1}$ [see Eq. (\ref{Rudiger})], we can calculate $D_{1}$ in terms of $D_0$ as:  
\EQ
D_{1} = \frac{1}{D_0} \int_{1/l_{1}}^{1/l_0}  P(k,\nu,\Omega) \, dk \; ,
\EN 
where $P(k,\nu, \Omega)$ is a function depending only on modulus of the wave number, molecular viscosity, and rotation. Note that the integration range is over the scale $l$ such that $l_0 < l < l_1$. Repeating this procedure iteratively, one can compute the turbulent diffusivity at scale $n+1$  as follows:
\EQ
\label{BullShit180}
D_{n+1} = \frac{1}{D_0+D_n} \int_{1/l_{n+1}}^{1/l_n}  P(k,\nu,\Omega) \, dk \; .
\EN 
If we take the limit $n \rightarrow \infty$, we obtain a continuous range of scale and turbulent diffusivity. Thus, writing $D(k)$ as the turbulent diffusivity acting on scales smaller than $k^{-1}$, Eq. (\ref{BullShit180}) can be rewritten
\EQ
dD(k) = \frac{1}{D_0+D(k)} \, P(k) \, dk \; .
\EN
The turbulent diffusivity $D_T$ is obtained by integrating over all scales in the inertial range, starting from $n=0$ with the value of molecular diffusivity $D_0=D$, with the result:
\EQ
\label{BullShit190}
D_T \sim \left( \frac{1}{D}  \int_0^{\infty} P(k,\nu,\Omega) \, dk \right)^{1/2} \; .
\EN
Eq. (\ref{BullShit190}) shows that if the quasi-linear theory gives a turbulent diffusivity as $\Omega^{-n}$, then the renormalisation analysis predicts a scaling $\Omega^{-n/2}$. Consequently, as the quasi-linear theory predicted a scaling as $\Omega^{-1}$ (irrespective of the form of temporal correlations), the turbulent transport of particles (or heat) scales as $\Omega^{-1/2}$ in the large rotation limit. Interestingly, this result has a weaker dependence on $\Omega$ compared to that according to quasi-linear theory. Furthermore $D_T^{xx} / D_T^{zz} = \sqrt{2}$ in the rapid rotation limit, with a weaker anisotropy compared to the quasi-linear result [see Eq. (\ref{Relation})].

Finally, we note that quasi-linear or multi-scale renormalisation group analysis  cannot capture a strong anisotropy in the rapid rotation limit with a tendency towards two-dimensionalisation  \citep{Cambon97}. In order to capture the highly anisotropic nature of rotating turbulence, one needs to use more sophisticated closure, such as wave-turbulence which leads to anisotropic spectra \citep{Galtier03}. In the extreme limit of rapid rotation, this rotation-induced anisotropy could be captured by assuming geostrophic balance in the equilibrium state \citep{Pedlovsky}.

\section{Summary and discussion}
\label{Discussion}
We studied the structure of turbulence in a rotating medium with an arbitrary external forcing. Using the quasi-linear theory, we first performed a thorough study of the turbulence intensity and transport (of chemical species or heat) in the case of an isotropic forcing by considering different types of temporal correlation functions. We also examined the transport of angular momentum in the case of an anisotropic forcing.

Specifically, we showed that for an isotropic forcing, the transport parallel and perpendicular to the rotation vector  has the same scaling with the rotation frequency $\Omega$: they are both reduced by a factor $\Omega^{-1}$ compared to the case without rotation irrespective of the temporal correlation function of the forcing. Furthermore, the transport in the direction parallel to the rotation is twice as fast than the one in the perpendicular direction [see Eq. (\ref{Relation})]. This result was shown to be robust independent of the specific form of temporal correlation $C(\tau)$ (exponential and power law) or correlation time (finite, short or infinite). Note that this result was shown by \citet{Kichatinov94} in the case of an infinitely correlated forcing. 

In comparison, the scaling of the turbulence intensity depends on the property of the forcing. For exponential correlation function with infinite memory, all the components of the velocity amplitude scale as $\Omega^{-1}$  while their intensity in the direction parallel to the rotation is twice that in the perpendicular direction (similar to the transport). However, for $\delta$-correlated turbulence, the turbulence amplitude becomes independent of the rotation rate. Thus, in this case, inertial waves affect only the cross-phase of the velocity field and not its amplitude. Furthermore, in the case of a finite correlation time with a power-law, the turbulence intensity scales as $\Omega^n$ with $-1 < n < 0$ while the ratio of the turbulence amplitude in the direction parallel to that perpendicular to the rotation can be any number between $1$ and $2$ depending on the exponent of the power-law.   

In the case where the driving force is highly anisotropic, it was shown to give rise to non-diffusive fluxes when the rotation is perpendicular to the anisotropy. In the large rotation limit, the transport of angular momentum scales as $\Omega^{-2}$ for a forcing with a finite correlation time. However, for a $\delta$-correlated turbulence, we found that the (quenching) effect of rotation was reduced with non-diffusive fluxes now scaling as $\Omega^{-1}$. Furthermore, for a power-law correlation function, the momentum flux was found to scale as $\Omega^{p}$ with $-2 < p < -1$. 

Finally, we performed a multi-scale renormalisation analysis which permitted us to go beyond the quasi-linear theory in the calculation of the turbulent diffusivity. We found that, in comparison with the quasi-linear theory result of a turbulent diffusivity proportional to $\Omega^{-1}$, the renormalisation theory predicts a scaling $\Omega^{-1/2}$. Consequently, we expect the turbulent diffusivity to depend weakly on $\Omega$. It also gave us an anisotropy in transport weaker than the quasi-linear prediction, the parallel component being $\sqrt{2}$ faster than that of the perpendicular. 

We should however note that the scaling of our results can be modified in the case of a bounded domain of finite size. In the calculation of all the turbulent transport coefficients, we obtained a result proportional to the following type of integral:
\EQ
\label{Bullshit300}
I({\bf k},\Omega) = \int \frac{H(k)}{\nu^2 k^4 + \omega_0^2} \, d^3 k \; ,
\EN
where $\omega_0 = ({\bf \Omega \cdot k}) / k$ is the projection of the unit vector in the direction of the wave number on the rotation axis. When the domain of integration is unbounded, the integration over the angular variable makes this integral proportional to $\Omega^{-1}$, when the rotation rate $\Omega$ is sufficiently large [e.g. Eq. (\ref{Rudiger})]. This is because this integral involves some contribution of order $1$ (when ${\bf \Omega \cdot k} = 0$) and others of order $\Omega^{-2}$. However, in a realistic situation, the domain of integration (in Fourier space) is bounded. That is, there is a minimal wavenumber (corresponding to a maximum length, for instance the size of the box)  in the direction of the rotation, for example, $k_m = \mathrm{min}(k_x)$. The aforementioned scaling of $\Omega^{-1}$  is valid only when $\nu^2 k^6 \gg \Omega^2 k_m^2$. In the opposite case ($\nu^2 k^6 \ll \Omega^2 k_m^2$), the term  $\omega_0^2$ in Eq. (\ref{Bullshit300}) is always dominant, giving the scaling of $\Omega^{-2}$ for large rotation rate. In this case, the multi-scale analysis would give a scaling of $\Omega^{-1}$ instead of $\Omega^{-1/2}$.

\section{Implications for stellar convection zone}
\label{Implications}
We discuss here some of the important implications of our results for the Sun and other stars. For parameters typical of the Sun, $\Omega \sim 2.3 \times 10^{-6} \, \mbox{s}^{-1}$ and $\nu \sim 10^2 \, \mbox{cm}^2 \mbox{s}^{-1}$, the effect of a finite-size domain discussed in Sect. \ref{Discussion} can be ignored if the characteristic length scale is larger than $(\nu / \Omega)^{1/2} \sim 1.7 \times 10^4 \mbox{cm}$. This is likely to be the case in the solar convection zone as convective motions can extend to a distance of approximately $4 \times 10^9 \mbox{cm}$. In this case, the quasi-linear theory predicts that the turbulent diffusivity scales as  $\Omega^{-1}$ for rapid rotation while the turbulence amplitude scales as $\Omega^0-\Omega^{-1}$ depending on the property of the forcing. In comparison, the renormalisation analysis predicts a turbulent diffusivity scaling as $\Omega^{-1/2}$. This result thus suggests that the effect of rotation on particle or heat transport is weaker than previously thought \citep{Kichatinov94}. If the effect of rotation had been much stronger in reducing transport, it could have inhibited the mixing of light elements in the convection zone, thereby leading to a weak mixing as required \citep{Barnes99}. Our results have confirmed that this cannot possibly be true and that rotation cannot play an important role in reducing the mixing of light elements in the solar convection zone to explain the observed (rather modest) depletion of light elements on the Sun \citep{Schatzman91}.

Traditional stellar modelling does not involve any anisotropy in the turbulent diffusivity of particles (or equivalently in the turbulent conductivity of temperature) due to rotation. We have shown here that such an anisotropy can be induced by rotation. This is however not a significant effect as the transport in the direction parallel to the rotation is only twice as fast as that in the perpendicular direction, according to the quasi-linear theory. This anisotropy is quite robust against the temporal correlations used and has been shown to give rise to a warmer pole than the equator \citep{Kichatinov95}. This anisotropy is however shown to be weaker according to the multi-scale renormalisation analysis, the ratio being only $\sqrt{2}$. Consequently, rotation may not be  sufficient for inducing anisotropic transport in stars, for instance, in order to cause a sufficient latitudinal temperature gradient in the convective envelopes as required by numerical models \citep{Kichatinov95,Kuker05}. As we have shown in our previous studies, an anisotropic transport can easily be induced by a strong shear layer \citep[such as the solar tachocline, see][]{Kim05,2Shears}, magnetic fields \citep{Kim06,BetaPlane,Dynamics} or stratification \citep{Stratification}. In these works, shear flows are also shown to quench turbulent transport significantly leading to weak mixing.

Finally, we note that our results depend on the magnitude of $\Omega$, the diffusion rate $\nu k^2$ (or $D k^2$) and the characteristic frequency of the forcing $1/\tau_f$. Obviously, stars of different spectral types have different values for $\Omega$, $\nu$, $D$ and $\tau_f$. It is thus important to understand the origin of $\tau_f$ and to perform a systematic study  on the development of turbulence theory as a function of masses and ages. This will be addressed in future studies.

\begin{acknowledgements}
This work was supported by U.K. PPARC Grant No. PP/B501512/1.
\end{acknowledgements}

\begin{appendix}
\section{Turbulence amplitude and transport}
\label{GrossesExpr}
In this appendix, we provide the form of $A^i_{jk}(k,\omega_0)$ and $M^i_{jk}(k,\omega_0)$ used in Eqs. (\ref{TurbInt}) and (\ref{TranspAngular}).
\EQA
\label{BullShit150}
A^x_{11} &=& \left( \frac{1}{2\nu k^2} + \frac{\nu k^2}{2(\nu^2 k^4 + \omega_0^2)} \right) \kappa_c - \frac{\omega_0}{2(\nu^2 k^4 + \omega_0^2)} \sigma_c \; ,  \\ \nonumber
A^x_{12} &=&  - \theta \left[\frac{\omega_0}{2(\nu^2 k^4 + \omega_0^2)} \kappa_c + \frac{\nu k^2}{2(\nu^2 k^4 + \omega_0^2)} \sigma_c \right] \; ,\\ \nonumber
A^x_{22} &=& \left( \frac{1}{2\nu k^2} - \frac{\nu k^2}{2(\nu^2 k^4 + \omega_0^2)} \right) \kappa_c + \frac{\omega_0}{2(\nu^2 k^4 + \omega_0^2)} \sigma_c \; , \\ \nonumber
A^y_{11} &=& \Bigl(\frac{\gamma(\beta^2+a^2)}{2\nu k^2 a^2} + \frac{-\gamma(\beta^2+a^2)+2a^2}{a^2} \frac{\nu k^2}{2(\nu^2 k^4 + \omega_0^2)} \\ \nonumber
&& + \frac{2 \theta \beta \sqrt{\gamma+a^2}}{a}  \frac{\nu k^2}{2(\nu^2 k^4 + \omega_0^2)}\Bigr) \kappa_c  \\ \nonumber
&& +  \left(\frac{\gamma(\beta^2+a^2)}{a^2} \frac{\omega_0}{2(\nu^2 k^4 + \omega_0^2)} + \frac{ \theta \beta \sqrt{\gamma+a^2}}{a} \frac{\nu k^2}{\nu^2 k^4 + \omega_0^2}\right) \sigma_c  \; , \\ \nonumber 
A^y_{12} &=& \Bigl(\theta \frac{\gamma(\beta^2+a^2)-2a^2}{a^2} \frac{\omega_0}{2(\nu^2 k^4 + \omega_0^2)} \\ \nonumber 
&& + \frac{2 \beta \sqrt{\gamma+a^2}}{a}  \frac{\nu k^2}{2(\nu^2 k^4 + \omega_0^2)}\Bigr) \kappa_c \\ \nonumber 
&&  -  \Bigl(\theta \frac{\gamma(\beta^2+a^2)-2a^2}{a^2} \frac{\nu k^2}{2(\nu^2 k^4 + \omega_0^2)} \\ \nonumber
&& - \frac{2 \beta \sqrt{\gamma+a^2}}{a}  \frac{\omega_0}{2(\nu^2 k^4 + \omega_0^2)}\Bigr)\sigma_c \; , \\ \nonumber
A^y_{22} &=& \Bigl(\frac{\gamma(\beta^2+a^2)}{2\nu k^2 a^2} + \frac{\gamma(\beta^2+a^2)-2a^2}{a^2}\frac{\nu k^2}{2(\nu^2 k^4 + \omega_0^2)} \\ \nonumber
&&  - \frac{2 \theta \beta \sqrt{\gamma+a^2}}{a} \frac{\nu k^2}{2(\nu^2 k^4 + \omega_0^2)} \Bigr) \kappa_c  \\ \nonumber
&&  +  \Bigl(\frac{-\gamma(\beta^2+a^2)+2a^2}{a^2} \frac{\omega_0}{2(\nu^2 k^4 + \omega_0^2)} \\ \nonumber
&& - \frac{\theta \beta \sqrt{\gamma+a^2}}{a} \frac{\nu k^2}{\nu^2 k^4 + \omega_0^2}\Bigr)\sigma_c  \; , \\ \nonumber
M^x_{11} &=& \left(-\frac{1}{2\nu k^2} - \frac{\nu k^2}{2(\nu^2 k^4 + \omega_0^2)}  - \frac{\beta \theta \sqrt{\gamma+a^2}}{a} \frac{\omega_0}{2(\nu^2 k^4 + \omega_0^2)} \right) \kappa_c \\ \nonumber
&&  + \left(\frac{\omega_0}{2(\nu^2 k^4 + \omega_0^2)}  - \frac{\beta \theta \sqrt{\gamma+a^2}}{a} \frac{\nu k^2}{2(\nu^2 k^4 + \omega_0^2)} \right)\sigma_c \; , \\ \nonumber
M^x_{12} &=&   \left(-\beta \frac{\sqrt{\gamma+a^2}}{a} + \theta \frac{\omega_0}{2(\nu^2 k^4 + \omega_0^2)} \right) \kappa_c  + \theta \frac{\nu k^2}{2(\nu^2 k^4 + \omega_0^2)} \sigma_c \; , \\ \nonumber
M^x_{22} &=& \left(-\frac{1}{2\nu k^2} + \frac{\nu k^2}{2(\nu^2 k^4 + \omega_0^2)}  + \frac{\beta \theta \sqrt{\gamma+a^2}}{a} \frac{\omega_0}{2(\nu^2 k^4 + \omega_0^2)} \right) \kappa_c \\ \nonumber
&& + \left(-\frac{\omega_0}{2(\nu^2 k^4 + \omega_0^2)}  + \frac{\beta \theta \sqrt{\gamma+a^2}}{a} \frac{\nu k^2}{2(\nu^2 k^4 + \omega_0^2)} \right)\sigma_c  \; , \\ \nonumber
M^z_{11} &=& \Bigl(-\frac{\beta \gamma}{2 a^2 \nu k^2} + \frac{\beta (\gamma+2 a ^2) \nu k^2}{2 a^2 (\nu^2 k^4 + \omega_0^2)} \\ \nonumber
&&  - \frac{(1-\beta^2) \theta \sqrt{\gamma+a^2}}{a} \frac{\omega_0}{2(\nu^2 k^4 + \omega_0^2)} \Bigr) \kappa_c \\ \nonumber
&& - \Bigl(\beta \frac{(\gamma+2a^2) \omega_0}{2a^2(\nu^2 k^4 + \omega_0^2)}  \\ \nonumber
&& + \frac{(1-\beta^2) \theta \sqrt{\gamma+a^2}}{a} \frac{\nu k^2}{2(\nu^2 k^4 + \omega_0^2)} \Bigr) \sigma_c \; , \\  \nonumber
M^z_{12} &=&   \Bigl( (\beta^2-1) \frac{\sqrt{\gamma+a^2}}{a}\frac{\nu k^2}{2(\nu^2 k^4 + \omega_0^2)}  \\ \nonumber
&& - \theta \beta \frac{\gamma+2a^2}{a^2}  \frac{\omega_0}{2(\nu^2 k^4 + \omega_0^2)} \Bigr) \kappa_c \\ \nonumber
&& + \Bigl( - \theta \beta \frac{\gamma+2a^2}{a^2} \frac{\nu k^2}{2(\nu^2 k^4 + \omega_0^2)}  \\ \nonumber
&& - (\beta^2-1) \frac{\sqrt{\gamma+a^2}}{a} \frac{\omega_0}{2(\nu^2 k^4 + \omega_0^2)} \Bigr) \sigma_c  \; , \\ \nonumber
M^z_{22} &=& \Bigl(-\frac{\beta \gamma}{2 a^2 \nu k^2} - \frac{\beta (\gamma+2 a ^2) \nu k^2}{2 a^2 (\nu^2 k^4 + \omega_0^2)} \\ \nonumber
&& + \frac{(1-\beta^2) \theta \sqrt{\gamma+a^2}}{a} \frac{\omega_0}{2(\nu^2 k^4 + \omega_0^2)} \Bigr) \kappa_c  \\ \nonumber
&& + \Bigl(\beta \frac{(\gamma+2a^2) \omega_0}{2a^2(\nu^2 k^4 + \omega_0^2)}  \\ \nonumber
&& + \frac{(1-\beta^2) \theta \sqrt{\gamma+a^2}}{a} \frac{\nu k^2}{2(\nu^2 k^4 + \omega_0^2)} \Bigr)\sigma_c  \; . 
\ENA
Here again, the functions $\kappa_c$, $\sigma_c$ and $\zeta_c$ characterising the influence of the temporal correlation of the forcing on the turbulence are defined by Eqs. (\ref{BullShit10}) and (\ref{BullShit11}).

\end{appendix}

\bibliographystyle{aa}
\bibliography{Bib_Rotation}

\end{document}